# Load-velocity-temperature relationship in frictional response of microscopic contacts

*Wengen Ouyang[1], Yao Cheng[2], Ming Ma[2,*], Michael Urbakh[1,*]*

[1] School of Chemistry and The Sackler Center for Computational Molecular and Materials Science, Tel Aviv University, Tel Aviv 69978, Israel

[2] State Key Laboratory of Tribology, Department of Mechanical Engineering and Center for Nano and Micro Mechanics, Tsinghua University, Beijing 100084, People's Republic of China

[*]Corresponding author.
E-mail address: urbakh@tauex.tau.ac.il (M. Urbakh).
E-mail address: maming16@tsinghua.edu.cn (M. Ma)

ABSTRACT

Frictional properties of interfaces with dynamic chemical bonds have been the subject of intensive experimental investigation and modelling, as it provides important insights into the molecular origin of the empirical rate and state laws, which have been highly successful in describing friction from nano to geophysical scales. Using previously developed theoretical approaches requires time-consuming simulations that are impractical for many realistic tribological systems. To solve this problem and set a framework for understanding microscopic mechanisms of friction at interfaces including multiple microscopic contacts, we developed an analytical approach for description of friction mediated by dynamical formation and rupture of microscopic interfacial contacts, which allows to calculate frictional properties on the time and length scales that are relevant to tribological experimental conditions. The model accounts for the presence of various types of contacts at the frictional interface and predicts novel dependencies of friction on sliding velocity, temperature and normal load, which are amenable to experimental observations. Our model predicts the velocity-temperature scaling, which relies on the interplay between the effects of shear and temperature on the rupture of interfacial contacts. The proposed scaling can be used to extrapolate the simulation results to a range of very low sliding velocities used in nanoscale friction experiments, which is still unreachable by simulations. For interfaces including two types of interfacial contacts with distinct properties, our model predicts novel double-peaked dependencies of friction on temperature and velocity. Considering friction force microscopy experiments (FFM), we found that the non-uniform distribution of normal load across the interface leads to a distribution of barrier heights for contact formation. The results obtained in this case allowed to reveal a mechanism of nonlinear dependence of friction on normal load observed in recent FFM experiments and predict the effect of normal load on velocity and temperature dependencies of friction. Our work provides a promising avenue for the interpretation of the experimental data on friction at interfaces including microscopic contacts and open new pathways for the rational control of the frictional response.

Keywords: chemo-mechanical processes (A), friction (B), asymptotic analysis (C), multi-contact model

# 1. Introduction

Friction plays an important role in diverse systems ranging from nanoscale contacts [1, 2] and microscale biological structures [1, 3], to the earthquakes and faults at geophysical [4]. Although friction has been studied for centuries, predicting sliding friction between two solids from the fundamental principles is still impossible. The reason is that friction is a highly non-equilibrium processes and depends on both the intrinsic physical and chemical properties of interfaces, and on external conditions, such as normal load, sliding velocity, temperature and humidity. Due to the surface roughness, most macroscopic contacts between two solids include multiple asperities, and the measured static friction increases logarithmically with the contact time. This phenomenon is referred to as frictional aging, which is usually attributed to the increase of contact area due to plastic creep [5-7] or to the formation of capillary bridges between two surfaces [8-11]. However, recent experiments performed at well-designed interfaces showed that the frictional aging may occur primarily due to the formation of interfacial bonds, and not from plastic creep or capillary bridges [12, 13]. Apart from frictional aging, it was found that the interfaces with the capability to form chemical bonds exhibit novel dependencies of sliding friction on temperature and velocity [14], which cannot be explained by the widely used Prandtl-Tomlinson (PT) model [15]. To reveal microscopic mechanisms underlying the observed non-monotonic temperature and velocity dependencies of friction, a new conceptual model (multi-contact model (MCM)) [14, 16] was proposed, which describes friction in terms of thermally activated formation and rupture of contacts between two surfaces in relative motion. The simulated temperature and velocity dependencies of friction based on this model agree well with the experimental observations [14].

An important advantage of this conceptual approach is that it can be used to describe various types of interfacial contacts characterized at different length and time scales. The contacts defined in this model can mimic molecular bonds [14, 17, 18], macromolecular complexes [19], capillary bridges [8-10], adhesion junctions in muscles and cells [20-22] and asperities at rough surfaces [23-29]. However, until now applications of this approach are limited, since they require performing time-consuming numerical simulations, which become impractical for many realistic tribological systems. To overcome this problem, one has to derive an analytical equation for the friction force, which provides a good approximation for the simulation results. First step in this direction has been done in Refs. [20-22]. However, the derived equations don't include effects of normal load and temperature that are the key parameters in the tribological phenomena, and consider only one type of interfacial contacts, that is usually not the case for realistic interfaces. Moreover, these equations cannot explain novel frictional behaviors observed in recent experiments [14, 30, 31], see details in Sec. 3.6.

The field of tribology and especially of nanotribology suffers from a lack of analytical equations that allow the analysis of experimental data at time and length scales that are relevant to measurements, and which can serve as a guide for future experimental studies and predict new phenomena. In this paper, we develop an analytical approach for consideration of frictional response of microscopic contacts, which extends the results of previous works [18, 22], including effects of normal load and temperature, as well as the presence of various types of contacts, which are responsible for multiple dissipative mechanisms that contribute to the overall friction. Analytical equations for the steady-state friction derived for various distributions of contact properties predict novel dependencies of friction on velocity, temperature, and normal load, which are amenable to experimental observations [14, 30, 31]. Thus, our work provides a promising avenue for description of friction at interfaces with multiple microscopic contacts.

## 2. Multi-contact model for a single type of interfacial contacts

### 2.1. Theoretical framework

The multi-contact model investigated in this paper is depicted in **Fig. 1**. The model includes two rigid surfaces connected by contacts ($N$ of them) that break under sliding and then reform upon contacting. The top surface (slider) moves with velocity $V(t)$ over the bottom surface (substrate). We model the contacts by linear elastic springs, each with a force constant $\kappa_b$. For simplicity, we assume that the contacts are present in one of two states: bound (B) or unbound (U). As long as a contact is in the bound state (unbroken), it is stretched in the lateral direction during sliding, while a ruptured contact relaxes rapidly to its unbound state. The rates of formation and rupture of contacts connecting two surfaces in relative motion are denoted as $k_{\text{on}}$ and $k_{\text{off}}$, respectively. Note that the two-state description of contacts is justified when the relaxation rate is higher than the rates of contact formation and rupture.

It should be noted that the meaning of the term "multi-contact" used here differs from the term "multi-asperity" commonly used for describing contacts between rough surfaces. Here, it refers to multiple contacts (or junctions) formed at the interface, which can mimic molecular bonds, macromolecular complexes, capillary bridges or adhesion junctions in muscles and cells [15]."

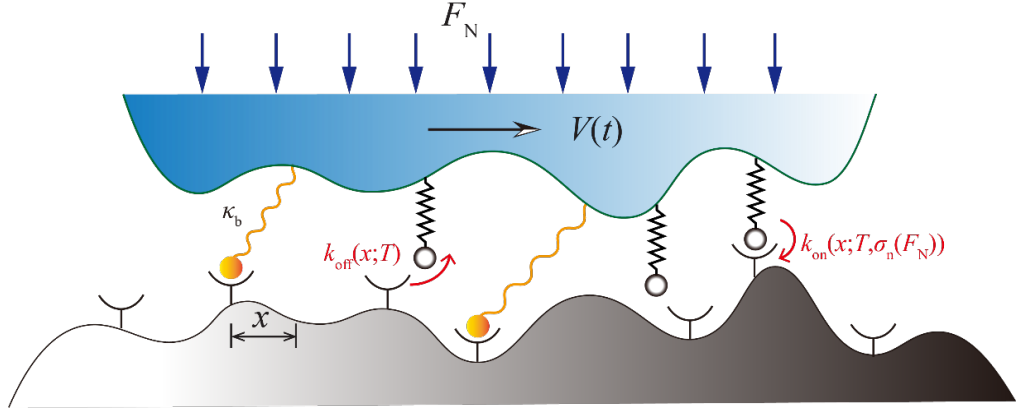

**Fig. 1**. Schematic of the multi-contact model. A slider moves with velocity $V(t)$ over a substrate covered with randomly distributed molecular contacts characterized by the spring constant, $\kappa_b$, the binding and unbinding rates of $k_{\text{on}}$ and $k_{\text{off}}$, respectively. Bound contact is stretched by the moving slider, and the force acting on it is $f = \kappa_b x$, where $x$ is the elongation of the contact. $F_N$ is the normal load applied to the slider.

In the limit of dense binding sites, the multi-contact model can be described by a so-called Lacker-Peskin partial differential equation (PDE) derived by [18]:

$$\frac{\partial \phi_b(x,t,\sigma_n)}{\partial t} + V(t)\frac{\partial \phi_b(x,t,\sigma_n)}{\partial x} = [1 - p_b(t,\sigma_n)]k_{\text{on}}(x;T,\sigma_n) - \phi_b(x,t,\sigma_n)k_{\text{off}}(x;T) \quad ,$$

(1)

where $\phi_b(x,t,\sigma_n)$ is the probability density function (PDF) to find a bound contact with elongation between $x$ and $x+dx$ at time $t$ under the contact pressure $\sigma_n$. Then, the probability that a contact is in the bound state at time $t$ is given by $p_b(t,\sigma_n) = \int_0^\infty \phi_b(x,t,\sigma_n)dx$. Here, the elongation $x$ is defined only for bound contacts [18] and as a result, $\phi_b(x<0,t,\sigma_n) = 0$. The rupture and formation of bound contacts are thermally activated processes, and their rates $k_{\text{on}}(x;T,\sigma_n)$ and $k_{\text{off}}(x;T)$, which depend on the contact elongation, temperature and pressure, can be written as follows [18, 22]:

$$\begin{cases} k_{on} = k_{on}^1 \delta(x), & k_{on}^1 = k_{on}^0 \exp(-\Delta E_{on}/k_B T) \\ k_{off}(x;T) = k_{off}^1 e^{ax}, & k_{off}^1 = k_{off}^0 \exp(-\Delta E_{off}/k_B T) \end{cases} \quad (2)$$

where $a = \kappa_b x_s / k_B T$, $x_s$ is the minimum-to-barrier distance in the potential describing contact-surface interaction [32, 33], $k_{on}^0, k_{off}^0$ and $\Delta E_{on}, \Delta E_{off}$ are the attempting frequencies and the barrier heights for the contact formation and rupture, respectively, $V(t)$ is the sliding velocity of the slider and $T$ is the temperature. In Eq. 2, the delta function $\delta(x)$ accounts for the assumption that ruptured contacts relax rapidly to their equilibrium state, which serves as an initial state for a formation of new contacts. There is a possibility of distribution of initial states for contact formation, and in this case, the delta function $\delta(x)$ should be replaced by a corresponding distribution function, for instance by a Gaussian function, as it was assumed by Qian et al [34, 35]. The applied load influences the kinetics of contact formation through its effect on the height of energy barrier for contact formation, $\Delta E_{on}$. This contribution to $\Delta E_{on}$ can be described by the equation [36-38],

$$\Delta E_{on} = \Delta E_{on}^0 - \sigma_n V_a, \quad (3)$$

where $\Delta E_{on}^0$ is the barrier height for contact formation in the absence of contact pressure, and $V_a$ is a parameter that is usually interpreted as the activation volume [38, 39]. Thus, the kinetic of bond formation may be influenced by the distribution of contact pressure across the interface, $\sigma_n(r, F_N)$, where $r$ is a coordinate along the interface.

In the presented model, the frictional dynamics is described by a one-dimensional equation of motion, considering the forces acting in the sliding (tangential) direction. However, the equilibrium state of the interface depends on the normal load, as, for instance described by Eqs. (11) and (42). The normal load significantly influences the frictional dynamics through its effect on the heights of activation energy barriers and on the actual contact area. It should be noted that deformations of the slider and substrate are not considered explicitly here, although their effect on the formation of the chemical bonds are taken into account effectively by introducing a normal load dependence of activation energy barriers. This simplification is justified for nanoscale contacts and materials with high Young's modulus [15]. The in-plane deformations may be important at larger scales for materials with low Young's modulus, where their effect on the deformation of the chemical bonds may be significant, as discussed by Qian et al [34, 35]. The extension of our model, including the coupling between the shear stress and normal stress and the deformation of the slider and substrate, is left for further works.

We can simplify Eq. (1) by integrating it over $x$ from $x = -\epsilon$ to $x = \epsilon$ ($\epsilon > 0$) that gives:

$$\left.\frac{\partial \phi_b(x,t,\sigma_n)}{\partial t}\right|_{-\epsilon}^{+\epsilon} + V[\phi_b(+\epsilon, t, \sigma_n) - \phi_b(-\epsilon, t, \sigma_n)] = \left[1 - \int_{-\infty}^{\infty} \phi_b(x, t, \sigma_n)dx\right]k_{on}^1 \int_{-\epsilon}^{+\epsilon} \delta(x)dx - \int_{-\epsilon}^{+\epsilon} \phi_b(x, t, \sigma_n)k_{off}(x)dx.$$

By setting $\epsilon \to 0^+$ and considering that $\phi_b(x < 0, t, \sigma_n) = 0$, we have

$$V\phi_b(0^+, t, \sigma_n) = \left[1 - \int_0^{\infty} \phi_b(x, t, \sigma_n)dx\right]k_{on}^1. \tag{4}$$

Since $\delta(x \neq 0) = 0$, Eq. (1) can be reduced for $x > 0$ to:

$$\frac{\partial \phi_b(x,t,\sigma_n)}{\partial t} + V(t)\frac{\partial \phi_b(x,t,\sigma_n)}{\partial x} = -\phi_b(x, t, \sigma_n)k_{off}(x; T). \tag{5}$$

To obtain Eq. (4), we assume that all detached contacts relax to their equilibrium immediately, i.e., the relaxation rate is much faster than the binding and unbinding rates, as explained above. Then, the PDF, $\phi_b(x, t, \sigma_n)$, can be found solving Eq. (5) with the boundary condition given by Eq. (4). These equations can be used to study the effect of bond formation and rupture on static friction and onset of sliding [10], but derivation of analytical equations describing these phenomena requires additional approximations compared to those used for consideration of steady-state friction. In the steady-state regime of motion the PDF does not change with time, i.e., $\phi_b(x, t, \sigma_n) = \phi_b(x, \sigma_n)$, and Eqs. (4) and (5) are reduced to

$$\begin{cases} V\frac{\partial \phi_b(x,\sigma_n)}{\partial x} = -\phi_b(x, \sigma_n)k_{off}(x) \\ V\phi_b(0^+, \sigma_n) = \left[1 - \int_0^{\infty} \phi_b(x, \sigma_n)dx\right]k_{on}^1 \end{cases}. \tag{6}$$

The analytical solution of the above equations reads as

$$\phi_b(x, \sigma_n) = \frac{ar_{on}(\sigma_n)Be^B}{1+r_{on}(\sigma_n)Be^B E_1(B)}\exp(-Be^{ax}), r_{on}(\sigma_n) = \frac{k_{on}^1}{k_{off}^1}. \tag{7}$$

where $E_1(B) = \int_B^{\infty} \frac{e^{-t}}{t}dt$ is the exponential integral and

$$B = \frac{k_{off}^1}{aV} = \frac{k_B T}{\kappa_b x_s V}k_{off}^0\exp\left(-\frac{\Delta E_{off}}{k_B T}\right) \tag{8}$$

being the key parameter in our theoretical model.

## 2.2. Analytical solution of the model at steady-state

Using the above expression for the PDF, $\phi_b(x)$, the number of bound contacts and the friction force at the steady-state can be calculated as:

$$N_{ss}^b(\sigma_n) = \int_S \rho(\mathbf{r})d\mathbf{r} \int_0^\infty \phi(x,\sigma_n)dx, \tag{9}$$

$$F_{ss} = \int_S \rho(\mathbf{r})d\mathbf{r} \int_0^\infty \kappa_b x \phi(x,t)dx. \tag{10}$$

Below we present some general results obtained for the uniform contact pressure distribution, which is a case for contacts between flat surfaces, such as the experimental systems used in surface force apparatus or self-retraction experiments [1, 40, 41]. The effect of non-uniform distribution of contact pressure on friction is discussed in section 3.5 in the context of friction force microscopy (FFM) experiments.

For the uniform distribution of applied normal load, $\sigma_n = F_N/S$, where $S$ is the contact area. Here, we further assume that the binding sites are uniformly distributed across the interface, i.e., $\rho(\mathbf{r}) = n_0 = $ const., where $n_0$ is number of binding sites per unit area, and $N = \int_S \rho(\mathbf{r})d\mathbf{r} = n_0 S$ is the total number of binding sites. Then, substituting Eq. (7) into Eqs. (9) and (10), we get

$$\begin{cases} N_{ss}^b(V,T,F_N;\boldsymbol{\xi}) = N \frac{r_{on}(\sigma_n)Be^B E_1(B)}{1+r_{on}(\sigma_n)Be^B E_1(B)} \\ F_{ss}(V,T,F_N;\boldsymbol{\xi}) = N \frac{k_B T}{x_s} \frac{r_{on}(\sigma_n)Be^B Q(B)}{1+r_{on}(\sigma_n)Be^B E_1(B)} \end{cases}, \tag{11}$$

where, $\boldsymbol{\xi} = \{N, E_{on}, E_{off}, k_{on}, k_{off}, \kappa_b, x_s\}$ is a set of intrinsic parameters that define the properties of the contacts, and

$$Q(B) = a^2 \int_0^\infty x \exp(-Be^{ax})dx = \int_1^\infty \frac{\ln y}{y} \cdot e^{-By} dy. \tag{12}$$

Using a definition of the hypergeometric function, $Q(B)$ can be written as follows:

$$Q(B) = \frac{\pi^2}{12} + \frac{\gamma^2}{2} + \gamma \ln B + \frac{1}{2}(\ln B)^2 - BG([1,1,1],[2,2,2],-B),$$

where $G$ is the generalized hypergeometric function, also called the Barnes extended hypergeometric function [42] and $\gamma = 0.5772156\cdots$ is the Euler-Mascheroni constant. Then, Eq. (11) can be rewritten as

$$F_{ss}(V,T,F_N;\boldsymbol{\xi}) = \frac{k_B T}{x_s} \frac{Q(B)}{E_1(B)} N_{ss}^b(V,T,F_N;\boldsymbol{\xi}), \tag{13}$$

which indicates that the friction force is proportional to the average number of contacts in the bound state

Temperature and velocity dependencies of friction predicted by Eq. (11) are illustrated in **Fig. 2**. For all values of sliding velocity considered here we found that the friction force varies non-monotonically with temperature, increasing at low temperatures and decreasing for higher temperatures (see **Fig. 2**a). This behavior is clearly observed in the 2D contour map (**Fig. 2**d) presenting the friction force as a function of temperature and velocity. The maximum of the friction force shifts to higher temperatures with increasing velocity. As shown in **Fig. 2**b, the velocity dependence of friction shows a similar bell-shaped behavior. Both velocity and temperature dependencies of friction predicted by Eq. (11) are consistent with the experimental data [14]. It's important to note that the predictions of the analytical Eq. (11) are in excellent agreement with the results of stochastic numerical simulation, no matter the slider is driven through a spring [14] or moved with a constant velocity, $V$ (see section 1 in supplementary material (SM) for details). It should be noted that in this paper, we focus on consideration of mechanism of frictional dissipation induced by the rupture and formation of interfacial chemical bonds (ICBI friction) and don't consider other sources of friction. As a result, the friction forces presented in **Fig. 2**a-b vanish in both low-temperature (high velocity) and high-temperature (low velocity) regimes. This is due to the fact that in the first regime only few contacts can be formed simultaneously during sliding, whereas in the second regime the rupture of contacts is determined by the rate of spontaneous unbinding and a time-averaged fraction of the intact contacts is close to its equilibrium value.

In **Fig. 2**c, we also show the load dependence of friction calculated for different sliding velocity at $T = 100$ K. It can be seen from this figure that the friction force is independent of the normal load in the low- and high-load regimes, and it increases nonlinearly with $\sigma_n$ in the intermediate range of loads. This behavior becomes evident by noting that the friction force scales with load as $F_{ss} \sim \frac{C \exp(\sigma_n V_a/k_B T)}{1 + C \exp(\sigma_n V_a/k_B T)}$, where $C$ is a parameter independent of load.

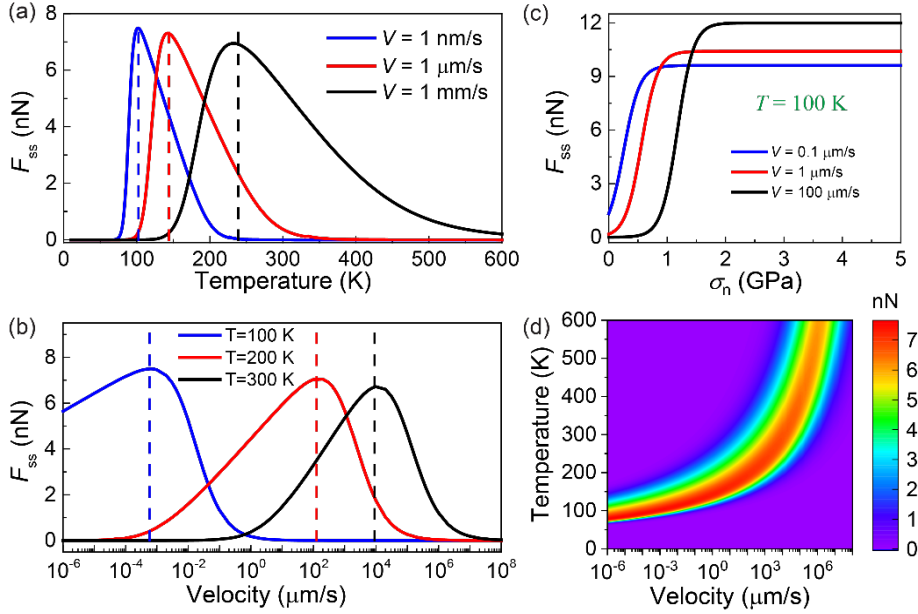

**Fig. 2**. Temperature, velocity and normal load dependencies of friction force calculated using Eq. (11). (a) Temperature dependence of friction for different sliding velocities at zero normal load. (b) Velocity dependence of friction for different temperatures at zero normal load. The dashed vertical lines indicated the positions of the peak temperatures and the velocities, which are calculated using Eqs. (18) and (16). (c) Normal load dependence of friction for different sliding velocities at $T = 100$ K. (d) Contour color map of friction as a function of temperature and velocity at zero normal load. Parameters used here: $k_{on}^0 = k_{off}^0 = 10^{13}$ s$^{-1}$, $\Delta E_{off} = 0.4$ eV, $\Delta E_{on} = 0.2$ eV, $\kappa_b = 10$ N/m, $x_s = 0.2$ nm, $N = 100$, $V_a = 10$ Å$^3$.

## 2.3. Asymptotic equations for the friction force

The advantage of the analytical Eq. (11) is its ability to predict frictional properties without time-consuming simulations, and provide quantitative insights about the dependence of friction on intrinsic properties of the system. For instance, Eq. (11) allows to derive expressions for the temperature and velocity that correspond to the maxima of the friction force as a function of $T$ and $V$, respectively. Considering that for most realistic systems, the height of energy barrier for contact rupture is significantly higher than that for the contact formation, i.e., $\Delta E_{off} - \Delta E_{on} \gg k_B T$, one can see that the parameter $r_{on}$ in Eq. (11) satisfies the inequality, $r_{on} \gg 1$. We found that under this condition the maxima of the friction force appear in the range of parameters, where $B \ll 1$, and the friction force can be as approximated by the equation

$$F_{ss}(V,T;\xi) = N \frac{k_B T}{2x_s} \frac{r_{on} B \ln^2 B}{1 - r_{on} B \ln B}. \tag{14}$$

Then, the positions of the maxima in the $(V, T)$ space can be found as solutions of equations $\partial F_{ss}(V,T;\xi)/\partial V \|_{V_m} = 0$ and $\partial F_{ss}(V,T;\xi)/\partial T \|_{T_m} = 0$. The first equation gives:

$$B(V_m) = \frac{1}{r_{on}}\left(1 + \frac{2}{\ln B(V_m)}\right) \approx \frac{1}{r_{on}}. \tag{15}$$

Substituting Eq. (8) into Eq. (15), we obtain the expression for velocity, $V_m$, corresponding to the maximum of the friction force as a function of $V$, which read as

$$V_m(T) = \frac{k_B T}{\kappa_b x_s} k_{on}^0 \exp\left(-\frac{\Delta E_{on}}{k_B T}\right). \tag{16}$$

The second equation, $\partial F_{ss}(V,T;\xi)/\partial T \|_{T_m} = 0$, gives:

$$B(T_m) = \frac{1}{r_{on}}\left(1 + \frac{1}{1+\lambda} + \frac{2}{\ln B(T_m)}\right) \approx \frac{1}{r_{on}}, \quad B(T_m) \ll 1, \tag{17}$$

where $\lambda = \Delta E_{off}/k_B T_m \gg 1$. Thus, the expression for temperature $T_m$ corresponding to the maximum of the friction force as a function of $T$ reads as

$$T_m(V) = \frac{\Delta E_{on}}{k_B W_0\left(\frac{k_{on}^0 \Delta E_{on}}{\kappa_b x_s V}\right)}. \tag{18}$$

where, $W_0(\cdot)$ is the principle branch of the Lambert $W$-function [43]. Eqs. (15) and (17) show that both $V_m$ and $T_m$ are determined by the same condition $B \approx \frac{1}{r_{on}}$.

Eqs. (16) and (18) predict that the peak velocity and temperature are determined by the parameters characterizing the contact formation, $k_{on}^0$ and $\Delta E_{on}$, and they are independent of the characteristics of the kinetics of contact rupture. In **Fig. 2**a-b the values of peak velocities and temperatures given by Eqs. (16) and (18) are marked by dashed vertical lines. One can see that they agree well with the results of exact calculation based on Eq. (11).

Besides predicting the positions of the peak velocity and peak temperature, analytical Eq. (11) allows to derive simple asymptotic expressions for the friction force, which are useful for the analysis of experimental results obtained under different external conditions. Interestingly, we found that the peak velocity, $V_m$, and the peak temperature, $T_m$, define the boundaries between regions, where the friction force exhibits different asymptotic behaviors as a function of velocity and temperature, respectively. For a given temperature, in the range of low and moderate velocities, $V < V_m$, where $B > 1/r_{on}$, the friction force can be approximated by the equation

$$F_{ss}(V,T;\xi) = N \frac{k_B T}{x_s} \frac{Q(B)}{E_1(B)} \tag{19}$$

whereas for high velocities, $V > V_\mathrm{m}$, where $B < 1/r_\mathrm{on}$, it can be described by Eq. (14). Correspondingly, Eqs. (14) and (19) approximate well the temperature dependence of friction for $T < T_\mathrm{m}$ and $T > T_\mathrm{m}$, respectively (see **Fig. 3**). It should be noted that approximations given by Eqs. (14) and (19) overlap in the range of moderate velocities ($V < V_\mathrm{m}$) and temperatures ($T > T_\mathrm{m}$).

Thus, our theory reveals the relationship between the velocity and temperature dependencies of friction. The frictional response in the range of low and moderate velocities corresponds to that for high temperatures, whereas the high velocity regime of friction corresponds to the low temperature one. In the context of friction, the hypothesis about velocity-temperature relationship was discussed previously [44, 45], assuming the possibility of shear-induced phase transitions in the interfacial layer. However, our theory provides a quantitative understanding of the mechanism of velocity-temperature relationship in nanoscale friction, and it relies on the interplay between the effects of shear and temperature on the rupture of interfacial contacts, rather than on the phase transition. Furthermore, Eq. (19) demonstrates that in the range of low and moderate velocities, usually considered in nanoscale friction experiments [15], the velocity and temperature dependencies of the friction force normalized by thermal energy, $F_\mathrm{ss}(V,T;\xi)/k_B T$, are determined by one scaling factor, $B = \frac{k_B T}{\kappa_b x_s V} k_\mathrm{off}^0 \exp\left(-\frac{\Delta E_\mathrm{off}}{k_B T}\right)$, which includes both intrinsic ($\Delta E_\mathrm{off}$, $k_\mathrm{off}^0$, $\kappa_b, x_s$) and external ($V,T$) parameters. Thus, the friction-velocity curves, measured for different temperatures, should collapse to a single master-curve, when they are plotted in the coordinates $F_\mathrm{ss}(V,T;\xi)/k_B T$ vs. $V/V^*$, where $V^* = \frac{k_B T}{\kappa_b x_s} k_\mathrm{off}^0 \exp\left(-\frac{\Delta E_\mathrm{off}}{k_B T}\right)$. The velocity-temperature scaling predicted here resembles the time-velocity superposition concept, which is widely used in polymer rheology [46-49], where it usually relies on the presence of phase transition. As we emphasized above, the mechanism of scaling derived from our model is quite different.

The proposed scaling can be used to compare experimental and simulation results at the overlapping velocities, which is still an unresolved problem in studies of nanoscale friction [15, 50, 51], since the molecular dynamics simulations become extremely time consuming and even inaccessible for the low sliding velocities used in measurements. To overcome this problem, the friction force can be calculated at a relatively high sliding velocities for different temperatures, which is less expensive in terms of simulations, and then the obtained results can be extrapolated to the range of low sliding velocities using the scaling relation.

Eq. (19) describing the friction force for $V < V_m$ can be further simplified considering the range of very low velocities, $V \ll V^*$, where $B \gg 1$, and moderate velocities, $V^* < V < V_m$, where $\frac{1}{r_{on}} < B < 1$. Corresponding asymptotic equations for the friction force read as

$$F_{ss} \sim N \frac{k_B T}{x_s} \frac{1}{B} = N \frac{\kappa_b x_s}{k_{off}^1} V \propto V, \qquad \text{for } V \ll V^*, \tag{20a}$$

$$F_{ss} \sim N \frac{k_B T}{2 x_s} (-\ln B) = N \frac{k_B T}{2 x_s} \ln(V/V^*) \propto \ln V, \qquad \text{for } V^* < V < V_m, \tag{20b}$$

Thus, analytical Eq. (19) describes both linear and logarithmic dependencies of friction, which are characteristic for "thermolubric" (close to equilibrium) and stick-slip regimes of motion, respectively [15, 52-54]. The low velocity asymptotic expression for the friction force, which is equivalent to that given by Eq. 20a, has been recently derived using the Jarzynski equality [55, 56] for the analysis of one-dimensional PT model of friction [57].

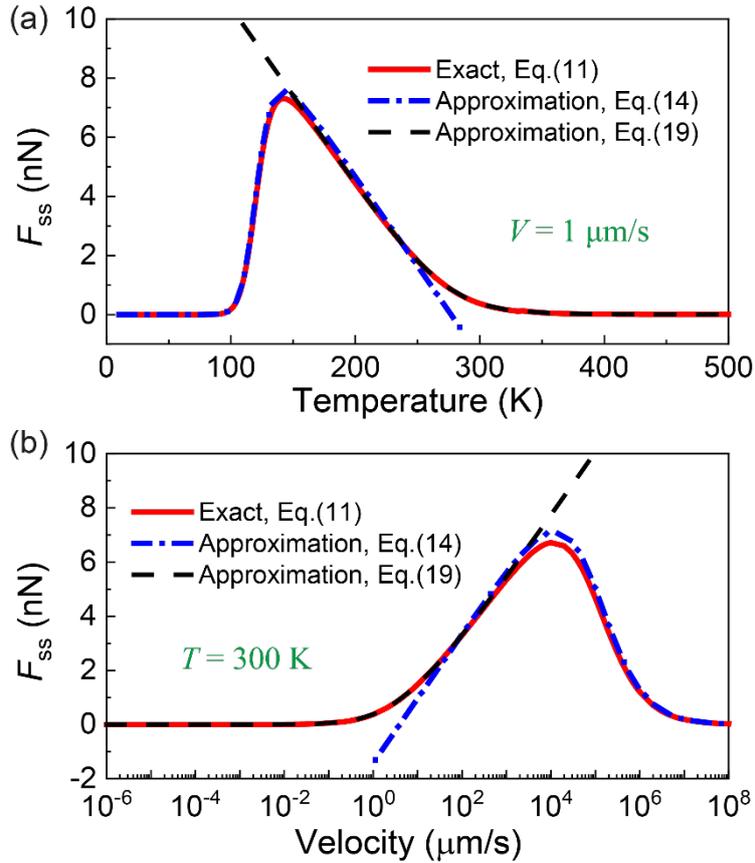

**Fig. 3**. Temperature (a) and velocity (b) dependencies of friction force calculated using Eq. (11) (red line) and its asymptotic expressions, Eq. (14) (blue dot-dashed line) and Eq. (19) (black dashed line), respectively. The sliding velocity in (a) is 1 μm/s and the temperature in (b) is 300 K. The parameters used here are the same as that in **Fig. 2**.

Similarly, Eq. (14) can be further simplified in the range of moderate velocities, $V^* < V < V_m$, where $\frac{1}{r_{on}} < B < 1$, and for very high velocities, where $V \gg V_m$ ($B \ll \frac{1}{r_{on}}$). The corresponding asymptotic equations read as

$$F_{ss} \sim N \frac{k_B T}{2x_s}(-\ln B) = N \frac{k_B T}{2x_s}\ln(V/V^*) \propto \ln V, \quad \text{for } V^* < V < V_m, \quad (21a)$$

$$F_{ss} \sim N \frac{k_B T}{2x_s} r_{on} B \ln^2 B \propto (\ln V)^2/V, \quad \text{for } V \gg V_m, \quad (21b)$$

Equations (20)-(21) demonstrate that the approximations for the friction force given by Eqs. (14) - (19) share the same asymptotic behavior in the range of moderate velocities. Moreover, our analysis shows that Eq. (14) provides a good description of the friction force for most values of velocity and temperature, except the range of very low $V$ and/or high $T$, which corresponds to the "thermolubric" regime of motion. A variety of different velocity regimes of friction and transitions between them, which are predicted by Eqs. (20)-(21), have been observed in various physical systems over a wide range of pulling velocities, including nanoscale friction [31], friction of optically trapped atoms [58] and mechanical unfolding of the muscle proteins [59].

# 3. Multi-contact model for a general distribution of contact properties

## 3.1. General properties

In the above section, we derived the analytical equation for the temperature, velocity and load dependencies of the steady-state friction force at interfaces including only one type of non-interacting contacts, characterized by unique values of barrier heights and attempting frequencies. However, recent experiments demonstrated that usually more than one type of contacts are present at a frictional interface [13, 30, 31], providing multiple dissipative mechanisms, which contribute to the overall friction. It should be noted that even for interfaces including only one type of contacts, an interaction between contacts or/and a non-uniform distribution of normal load across the interface, which is induced by surface roughness or is intrinsic for friction force microscopy (FFM) configuration, lead to a distribution of barrier heights for contact formation [10, 13, 19, 23, 31, 60-62]. In order to account for different types of interaction between sliding surfaces, we consider a general form of distribution of barrier heights for contact formation, whose probability density function (PDF) can be written as:

$$\tilde{\rho}(\Delta E_{on}) = \begin{cases} 0, & \Delta E_{on} \leq \Delta E_{on}^0 \\ \rho\left(\frac{\Delta E_{on} - \Delta E_{on}^0}{W}\right), & \Delta E_{on} > \Delta E_{on}^0 \end{cases}. \tag{22}$$

where, $\Delta E_{on}^0$ is the minimal barrier height and $W$ is the width of the distribution. The PDF, $\tilde{\rho}(\Delta E_{on})$, may depend on the normal load through its effect on the barrier heights, $\Delta E_{on}$. A detailed discussion of dependence of friction on normal load is provided in section 3.5.

Using the PDF described by Eq. (22) and neglecting interactions between contacts, the ensemble average steady-state friction force can be calculated as:

$$\overline{F_{ss}} = \int_{\Delta E_{on}^0}^{\infty} \rho\left(\frac{\Delta E_{on} - \Delta E_{on}^0}{W}\right) F_{ss}(V, T; \xi) d\Delta E_{on}. \tag{23}$$

Substituting Eq. (11) into Eq. (23), we get

$$\overline{F_{ss}} = \int_{\Delta E_{on}^0}^{\infty} \rho\left(\frac{\Delta E_{on} - \Delta E_{on}^0}{W}\right) \left[N \frac{k_B T}{x_s} \frac{Q(B)}{E_1(B)} \left(1 - \frac{1}{1 + Ar_{on}}\right)\right] d\Delta E_{on}. \tag{24}$$

where,

$$\begin{cases} A = Be^B E_1(B) \\ r_{on}(\Delta E_{on}) = \frac{k_{on}^0}{k_{off}^0} \exp[(\Delta E_{off} - \Delta E_{on})/k_B T] \end{cases}. \tag{25}$$

Defining $x = (\Delta E_{on} - \Delta E_{on}^0)/k_B T$, Eq. (24) can be rewritten as

$$\overline{F_{ss}} = N \frac{k_B T}{x_s} \frac{Q(B)}{E_1(B)} \left[1 - k_B T \int_0^{\infty} \frac{\rho\left(\frac{k_B T}{W} x\right)}{1 + Ar_{on}(\Delta E_{on}^0)e^{-x}} dx\right]. \tag{26}$$

Using Eq. (26), the average friction can be calculated once the distribution of $\Delta E_{on}$ is known. Following similar steps, we can calculate the average number of bound contacts as follows:

$$\overline{N_{ss}^b} = N \left[1 - k_B T \int_0^{\infty} \frac{\rho\left(\frac{k_B T}{W} x\right)}{1 + Ar_{on}(\Delta E_{on}^0)e^{-x}} dx\right]. \tag{27}$$

Comparing Eqs. (13), (26) and (27), we see that the relation between $\overline{F_{ss}}$ and $\overline{N_{ss}^b}$ derived for a single type of contacts, still works for contacts with a general distribution of energy barriers for the contact formation.

To make further progress in the evaluation of the friction force, we consider below three examples of PDFs of $\Delta E_{\text{on}}$, for which Eq. (26) can be integrated analytically. These PDFs, illustrated in **Fig. 4**, can be used to approximate realistic ones measured experimentally or obtained from simulation [61, 62]. Density functional theory (DFT) calculations presented by Liu *et al* [61] showed that there are multiple phenomena, which can control the distribution of energy barriers for formation and rupture of surface contacts, such as energetical and morphological inhomogeneity of frictional interface, and a distribution of pressures acting on surface atoms. In particular, it was found [61] the probability density function (PDF) of energy barrier heights for formation of bonds between contacting amorphous silica surfaces is highly non-uniform with a peak in the vicinity of 1 eV. The kinetic Monte Carlo (kMC) simulations [61] also demonstrated that interactions between the interfacial contacts broadens the PDF of barrier heights, and in a certain energy range the PDF becomes close to a uniform distribution. In addition, recent experimental studies [30, 31] indicated that several distinct types of contacts can be formed at frictional interfaces. In accordance with these observations, we consider below three typical PDFs for the barrier heights of bond formation at frictional interfaces:

i), uniform PDF, which accounts for the effect on surface morphology and contact interactions on the energy barriers;

ii), two (or multi) state PDF, which considers several distinct types of contacts formed at frictional interfaces;

iii), Gamma distribution, which captures the main feature of a realistic PDFs calculated for amorphous silica surfaces.

The first example is the uniform distribution (**Fig. 4**a), for which the PDF has the following form:

$$\tilde{\rho}(\Delta E_{\text{on}}) = \begin{cases} 0, \Delta E_{\text{on}} < \Delta E_{\text{on}}^1 \\ \frac{1}{W}, \quad \Delta E_{\text{on}}^1 \leq \Delta E_{\text{on}} \leq \Delta E_{\text{on}}^2, \\ 0, \quad \Delta E_{\text{on}} > \Delta E_{\text{on}}^2 \end{cases} \quad (28)$$

where $W = \Delta E_{\text{on}}^2 - \Delta E_{\text{on}}^1$ is the width of the uniform distribution.

The second example is the Gamma distribution (**Fig. 4**b) for which the PDF is given by the equation

$$\rho(\Delta E_{\text{on}}) = \begin{cases} 0, & \Delta E_{\text{on}} < \Delta E_{\text{on}}^0 \\ \frac{(\Delta E_{\text{on}} - \Delta E_{\text{on}}^0)^{k-1}}{\Gamma(k)W^k} \exp\left(-\frac{\Delta E_{\text{on}} - \Delta E_{\text{on}}^0}{W}\right), & \Delta E_{\text{on}} \geq \Delta E_{\text{on}}^0 \end{cases}. \quad (29)$$

As shown in **Fig. 4b**, depending on the value of parameter $k$, the Gamma distribution monotonically decrease with $\Delta E_{on}$ ($k \leq 1$), or exhibits a maximum as a function of $\Delta E_{on}$ ($k > 1$), thus covering a broad range of possible behaviors. PDFs showing a maximum as a function of $\Delta E_{on}$, similar to those presented in **Fig. 4b** for $k > 1$, have been predicted by molecular dynamics simulations of contacts between silica surfaces [61, 62]

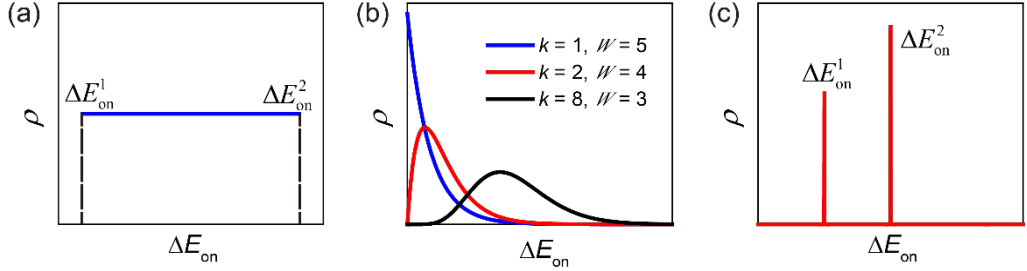

**Fig. 4.** Different types of PDFs of $\Delta E_{on}$: (a) a uniform distribution; (b) Gamma distribution; (c) two-state distribution.

The third example considered here is the two-state distribution, which could describe two types of contacts with distinct properties (see **Fig. 4c**). This PDF has the following form:

$$\tilde{\rho}(\Delta E_{on}) = c_1 \delta(\Delta E_{on} - \Delta E_{on}^1) + c_2 \delta(\Delta E_{on} - \Delta E_{on}^2). \tag{30}$$

The normalization of $\tilde{\rho}(\Delta E_{on})$ gives $c_1 = N_1/N$ and $c_2 = N_2/N$, where $N_1$ and $N_2$ are the numbers of available binding sites that correspond to two types of contacts with energy barrier heights of $\Delta E_{on}^1$ and $\Delta E_{on}^2$, respectively, and $N = N_1 + N_2$ is the total number of binding sites.

In the following sections, we derive analytical expressions for the steady state friction forces corresponding to the above PDFs.

## 3.2. Uniform PDF of energy barrier heights

In this case using Eqs. (26)-(27), the average number of bound contact and steady-state friction force can be calculated as follows:

$$\overline{N_{ss}^b} = \frac{1}{W} \int_{\Delta E_{on}^1}^{\Delta E_{on}^2} N \frac{r_{on} B e^B E_1(B)}{1 + r_{on} B e^B E_1(B)} d\Delta E_{on} = N \frac{k_B T}{W} \log \frac{1 + r_{on}^1 B e^B E_1(B)}{1 + r_{on}^2 B e^B E_1(B)} \tag{31}$$

$$\overline{F_{ss}} = \frac{1}{W} \int_{\Delta E_{on}^1}^{\Delta E_{on}^2} F_{ss}(V, T; \xi) d\Delta E_{on} = N \frac{k_B T}{W} \frac{k_B T}{x_s} \frac{Q(B)}{E_1(B)} \log \frac{1 + r_{on}^1 B e^B E_1(B)}{1 + r_{on}^2 B e^B E_1(B)} \tag{32}$$

where, $r_{on}^1 = \frac{k_{on}^0}{k_{off}^0} \exp\left(\frac{\Delta E_{off} - \Delta E_{on}^1}{k_B T}\right)$, $r_{on}^2 = r_{on}^1 \exp\left(\frac{W}{k_B T}\right)$.

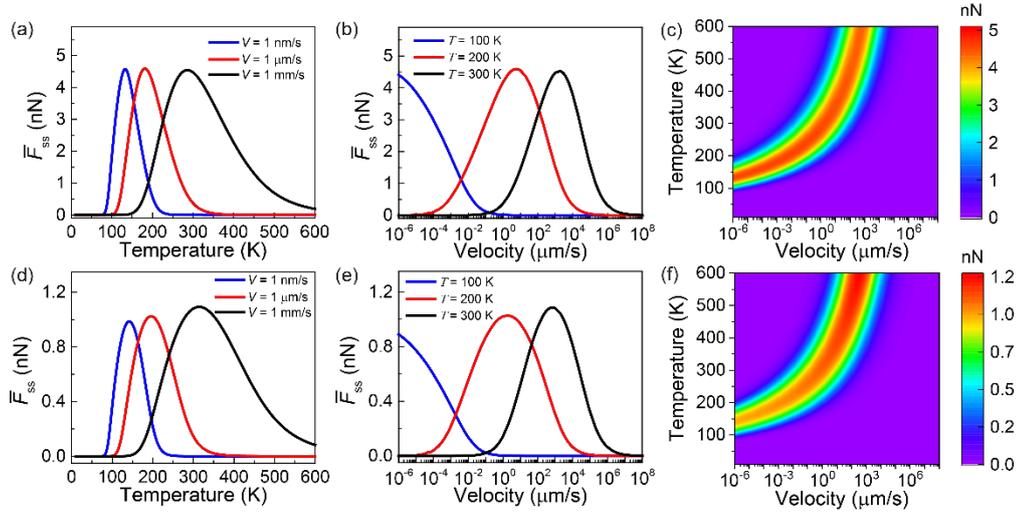

**Fig. 5.** Friction force as a function of temperature and velocity calculated for uniform PDFs with two different widths. The temperature (a, d), velocity (b, e) dependencies of steady-state friction and their 2D contour maps of friction in the (*T*, *V*) plane (c, f) for $W = 0.1$ eV and $W = 0.5$ eV, respectively. Parameter values used here: $k_{on}^0 = k_{off}^0 = 10^{13}$ s$^{-1}$, $\Delta E_{off} = 0.4$ eV, $\Delta E_{on} = 0.2$ eV, $\kappa_b = 10$ N/m, $x_s = 0.2$ nm, $N = 100$.

**Fig. 5** shows the velocity and temperature dependencies of friction force calculated for different widths of the uniform PDF. One can see that the variations of the friction force with *V* and *T* are similar to those calculated for a single type of contact (see **Fig. 2**), but the maximal friction force as functions of temperature and velocity decrease with increasing the width of the PDF of $\Delta E_{on}$. The latter is explained by the fact that the average value of $\Delta E_{on}$ increases as *W* increasing (under the condition that $\Delta E_{on}^1 =$ const), which makes it more difficult to form contacts, thus reduces the friction force. Note that the positions of the peaks in the $\overline{F_{ss}}$ vs *T* and *V* are almost independent of the PDF width.

*3.3. Gamma distribution*

The average steady-state friction force for the Gamma PDF of $\Delta E_{on}$ can be calculated as :

$$\overline{F_{ss}} = N \frac{k_B T}{x_s} \frac{Q(B)}{E_1(B)} \int_{\Delta E_{on}^0}^{\infty} \frac{(\Delta E_{on} - \Delta E_{on}^0)^{k-1}}{\Gamma(k) W^k} e^{-\frac{\Delta E_{on} - \Delta E_{on}^0}{W}} \cdot \frac{A r_{on}}{1 + A r_{on}} d\Delta E_{on}, \quad (33)$$

where $A = B e^B E_1(B)$ and $r_{on}(\Delta E_{on}) = \frac{k_{on}^0}{k_{off}^0} \exp[(\Delta E_{off} - \Delta E_{on})/k_B T]$. Setting $x = (\Delta E_{on} - \Delta E_{on}^0)/k_B T$, Eq. (33) can be simplified as

$$\overline{F_{ss}} = N \frac{k_B T}{x_s} \frac{Q(B)}{E_1(B)} \left[ 1 - \frac{\alpha^k}{\Gamma(k)} \int_0^\infty \frac{x^{k-1} e^{-\alpha x}}{1 + A r_{on}(\Delta E_{on}^0) e^{-x}} dx \right] = N \frac{k_B T}{x_s} \frac{Q(B)}{E_1(B)} [1 - \alpha^k \Phi(z, k, \alpha)] \quad , \quad (34)$$

where $z = -A r_{on}(\Delta E_{on}^0)$ and

$$\Phi(z, k, \alpha) = \frac{1}{\Gamma(k)} \int_0^\infty \frac{x^{k-1} e^{-\alpha x}}{1 - z e^{-x}} dx, \alpha = \frac{k_B T}{W} \quad (35)$$

is the Lerch transcendent function [63].

**Fig. 6** shows the velocity and temperature dependencies of friction for the gamma PDFs with different widths. Here we choose the PDFs for $k = 2$, exhibiting a maximum as a function of $\Delta E_{on}$. One can see that the variations of the friction force with $V$ and $T$ are similar to those presented in **Fig. 5**, namely, the maximal friction forces decrease with increasing the width of distribution. However, differently from the case of uniform PDF, here the positions of the friction peaks shift to higher temperatures and lower velocities with increasing the width of the distribution.

For $k = 1$, the Gamma distribution reduces to the exponential distribution, and in this case the PDF decreases monotonically with $\Delta E_{on}$ (see **Fig. 4**b). In section 2 of SM, we show the temperature and velocity dependencies of the friction force calculated for the exponential PDF with the same parameters as in **Fig. 6**. The qualitative behavior of the friction force as a function of temperature and velocity calculated for $k = 1$ is similar to that illustrated in **Fig. 6**. We found that with increasing $k$, the positions of the friction peaks shift to higher temperatures and lower velocities, respectively, and the magnitude of the friction force decreases significantly.

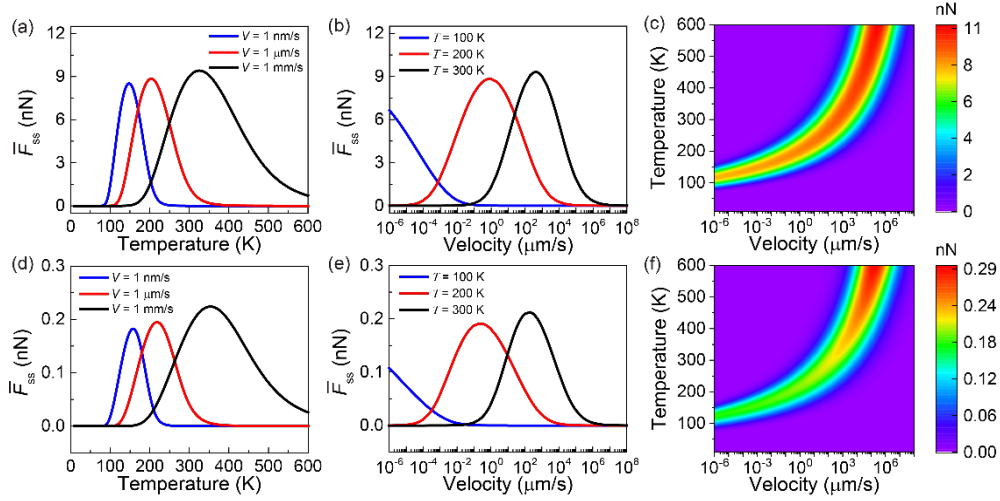

**Fig. 6.** Friction force as a function of temperature and velocity calculated for the Gamma PDFs with $k=2$ and two different widths. The temperature (a, d) and velocity (b, e) dependences of friction and the 2D contour maps of friction in $(T,V)$ plane (c, f) of steady-state friction for $W=0.1$ eV and $W=1$ eV, respectively. Parameter values: $k_{on}^0 = k_{off}^0 = 10^{13}$ s$^{-1}$, $\Delta E_{off} = 0.4$ eV, $\Delta E_{on} = 0.2$ eV, $\kappa_b = 10$ N/m, $x_s = 0.2$ nm, $N = 100$.

*3.4. Two-state distribution*

Recently, the first-principles calculations indicated that covalent interfacial siloxane (Si-O-Si) bridges are formed at silica-silica interfaces [61, 64]. FFM measurements of single-asperity silica-silica nanocontacts [30] found that two metastable states are involved in the process of formation and rupture of Si-O-Si bridges, which are the dangling state (Si-O-) and passivated state (Si-OH). Obviously, the heights of energy barrier for formation Si-O-Si bridges from these states are quite different. Therefore, the frictional response of this system can be described using the multi-contact model with two-state PDF of energy barriers, $\Delta E_{on}$. In this case the average friction force can be calculated as

$$\overline{F_{ss}}(V,T;\xi_1,\xi_2) = \frac{N_1}{N} F_{ss}(V,T;\xi_1) + \frac{N_2}{N} F_{ss}(V,T;\xi_2). \tag{36}$$

Here, $F_{ss}(V,T;\xi_1)$ and $F_{ss}(V,T;\xi_2)$ are the friction forces for the two distinct types of contacts, which can be calculated using Eq. (11). Calculations using Eq. (36) show that both the temperature and velocity dependencies of the average friction force may exhibit a "double-peak" behavior, illustrated in **Fig. 7**a-b, which differs significantly from those discussed above (see **Fig. 2**, 5, 6). The double-peak behavior is observed only in the case, where a difference in the heights of energy barriers for formation of two different contact is much larger than thermal energy, i.e., $\Delta E_{on}^2 - \Delta E_{on}^1 \gg k_B T$. This can be rationalized considering the effect of sliding velocity and temperature on the state of interfacial contacts. For $\Delta E_{on}^2 - \Delta E_{on}^1 \gg k_B T$, in a broad range of sliding velocities the contacts with lower energy barrier ($\Delta E_{on} = \Delta E_{on}^1$) are in the bound state, whereas the contacts with higher barrier height ($\Delta E_{on} = \Delta E_{on}^2$) can form bound states only for low enough sliding velocities. Thus, for low velocities, both types of contacts connect two surfaces, and in this regime the friction force increases with $V$, since the shear induced stretching of contacts increases with velocity. With further increase of velocity, an average number of contacts with higher energy barriers ($\Delta E_{on}$), which are present in the bound state, start to decrease, and as a result, the friction force decreases with $V$ reaching the minimum for velocity corresponding to the state, where all the contacts with higher $\Delta E_{on}$ ruptured. In this range of velocities the contacts with lower $\Delta E_{on}$ still can form the bound state, thus after reaching the minimum the friction force increases with $V$. Finally, in the high velocity regime, the average number of contacts with lower $\Delta E_{on}$ presenting in the bound state also starts to decrease leading to reduction of friction. The double-peak temperature dependence of friction shown in **Fig. 7**a can be understood in a similar way. Under the condition that the friction enhancement due to the shear-induced stretching of the contacts with lower $\Delta E_{on}$ is compensated by the decrease of friction caused by the reduction of number of contacts with higher $\Delta E_{on}$, presenting in the bound state, we found a plateau in the velocity and temperature dependencies of friction, as illustrated in **Fig. 7**c-d. The double-peak temperature dependence of friction has been observed in FFM experiments with a silicon tip sliding on a Si(111) and SiC wafer [14]. The double-peak temperature and velocity dependences of friction are clearly exhibited by the contour color maps in **Fig. 7**c, f. This effect can be explained by our two-state model (see details in Sec. 3.6). The two-state PDF considered here can be extended to the *N*-state distribution, for which the temperature and velocity dependence of friction may exhibit "*N*-peak" behavior.

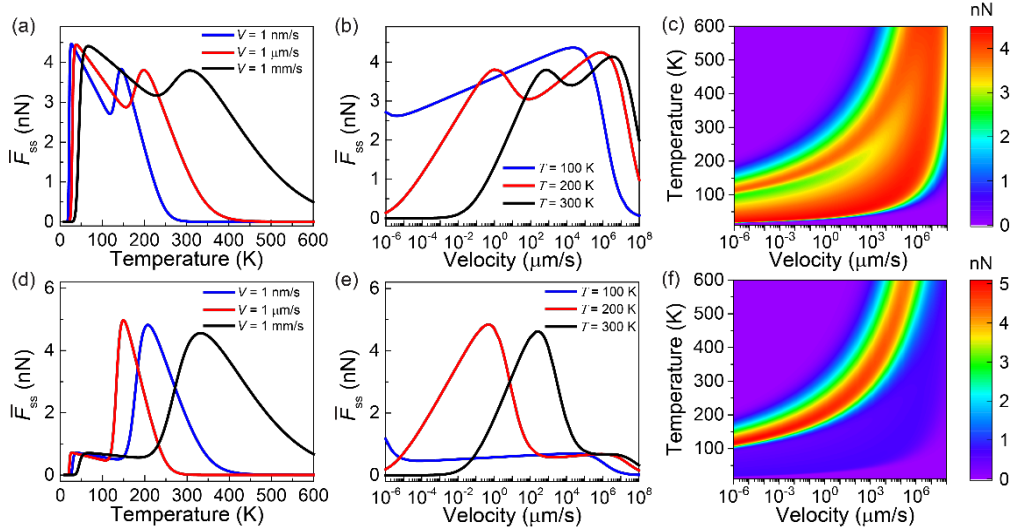

**Fig. 7.** Friction force as a function of temperature (a, c), velocity (b, e) and the 2D contour maps of friction in ($T,V$) plane (c, f) calculated for two-state PDFs. The results in panels (a, b, c) and (d, e, f) were obtained for $N_1 = 50$, $N_2 = 50$ and $N_1 = 20$, $N_2 = 80$, respectively. Parameter values used here: $k_{on}^0 = k_{off}^0 = 10^{13}$ s$^{-1}$, $\Delta E_{off} = 0.5$ eV, $\Delta E_{on}^1 = 0.05$ eV, $\Delta E_{on}^2 = 0.3$ eV, $\kappa_b = 10$ N/m, $x_s = 0.2$ nm.

*3.5. Multi contact model for FFM configuration*

Recent FFM experiments and simulations showed that nanoscale silica contacts exhibit aging due to the progressive formation of interfacial chemical bonds [12, 13, 23, 61, 62]. The role of normal load (or equivalently, normal stress) on this interfacial chemical bond-induced (ICBI) friction is predicted to be significant and recently has been examined experimentally [13] and numerically [23, 62]. However, an analytical model describing the mechanism of this phenomenon is still lacking.

In section 2.1, we presented a general formalism, which allows to consider the effect of normal load on the friction force and derived the analytic equations for the friction force as a function of velocity, temperature and normal load, which can be used for contacts between flat surfaces. However, almost all the experimental measurements of ICBI friction were performed using FFM. The typical configuration in FFM experiments is a AFM tip sliding on a flat surface. To study the frictional properties in this typical configuration, our model should be extended.

In this typical configuration, the application of load leads to two effects: (i) an enlargement of the contact area, which translates into an increase of the number of reaction sites available for contact formation, and (ii) a change in the height of energy barrier for the contact formation, which depends on the position of the contact with respect to the FFM tip apex. To include both the effects into our extended MCM, we assume that the load induced surface deformation of the contact can be described by the Hertzian contact theory [65] or by Derjaguin-Müller-Toporov (DMT) theory [66]. The latter takes into account the effect of adhesion, which plays an important role in many FFM experiments. Noting that Hertz theory is a special case of DMT theory, in the following section we derive the analytical equation for the friction force using the DMT theory.

Let's consider a spherical tip with radius $R_{\text{tip}}$, to which the normal force, $F_N$, is applied, then according to the DMT theory, the normal stress distribution is given by the equation [66]:

$$\begin{cases} \sigma_n(r; F_N) = \sigma_0(F_N)\left(1 - \frac{r^2}{a^2}\right)^{1/2} \\ a = \left[\frac{3R}{4E^*}(F_N + F_{\text{adh}})\right]^{1/3} \end{cases}, \tag{37}$$

where $r$ is the lateral distance from the tip apex, $\sigma_0(F_N) = \frac{3(F_N + F_{\text{adh}})}{2\pi a^2}$ is the normal stress at the center of the contact area, $a$ is the radius of contact area, $F_N$ is the external normal force applied to the tip and $F_{\text{adh}}$ is the adhesion force between the tip and the surface, $E^*$ is the effective Young's modulus of the contact: $\frac{1}{E^*} = \frac{1-\nu_1^2}{E_1} + \frac{1-\nu_2^2}{E_2}$, $\nu_1$ and $\nu_2$ are the Poisson's ratio of the tip and surface, respectively.

It should be noted that the proposed model neglects the coupling between the normal and tangential stress distributions, the consideration of which is beyond the scope of the Hertz model. The influence of shear stress on the normal stress distribution is characterized by the following parameter [65, 67]

$$\beta = \frac{1}{2}\left(\frac{1-2\nu_1}{G_1} - \frac{1-2\nu_2}{G_2}\right) / \left(\frac{1-\nu_1}{G_1} + \frac{1-\nu_2}{G_2}\right),$$

where $G_1$, $G_2$ and $\nu_1$, $\nu_2$ are the shear moduli and Poisson's ratios of the two intact elastic bodies, respectively. For contacts between identical materials or between uncompressible materials ($\nu = 0.5$), the parameter $\beta = 0$, and the Hertzian contact theory is exact. For typical tribological material pairs such as silicon/metal, silicon/graphite and graphite/metal [68-71], the parameter $\beta < 0.2$, and the Hertzian contact theory provides a very good approximation. Therefore, neglecting the coupling between the normal and tangential stress distributions is well justified for the tribological contacts considered in here. A more rigorous treatment of this issues is left for further work.

According to Eqs. (2)-(3), the rate of contact formation in this case reads as

$$k_{\text{on}}(r) = k_{\text{on}}^0 \cdot \exp(-\Delta E_{\text{on}}^0/k_B T) \cdot \exp[V_a \cdot \sigma_n(r; F_N)/k_B T]. \tag{38}$$

For a uniform distribution of contacts across the contact area, their radial density distribution can be written as:

$$\rho(r) = n_0 \cdot 2\pi r, \quad 0 \leq r \leq a, \tag{39}$$

where $n_0$ is the number of contacts per unit area, and the total number of contacts is $N = n_0 \cdot \pi a^2$. Under the above conditions, we can calculate the averaged friction force as follows:

$$\bar{F}_{ss} = \int_0^a dr \rho(r) \frac{k_B T}{x_s} \frac{Q(B)}{E_1(B)} \frac{A r_{\text{on}}(\sigma_n)}{1 + A r_{\text{on}}(\sigma_n)}, \quad A = B e^B E_1(B). \tag{40}$$

where, $r_{\text{on}}(\sigma_n) = r_{\text{on}}^0 \exp\left[V_a \cdot \frac{\sigma_n(r; F_N)}{k_B T}\right]$, $r_{\text{on}}^0 = \frac{k_{\text{on}}^0}{k_{\text{off}}^0} \cdot \exp\left(\frac{\Delta E_{\text{off}} - \Delta E_{\text{on}}^0}{k_B T}\right)$. Introducing parameter $x = V_a \cdot \frac{\sigma_n(r; F_N)}{k_B T}$, Eq. (40) can be rewritten as follows:

$$\bar{F}_{ss} = \frac{2n_0 \pi a^2}{x_0^2} \frac{k_B T}{x_s} \frac{Q(B)}{E_1(B)} \int_0^{x_0} dx \frac{xDe^x}{1+De^x}, \tag{41}$$

where $x_0 = \sigma_0(F_N) V_a/k_B T$ and $D = r_{\text{on}}^0 B e^B E_1(B)$. The integral in Eq. (41) can be written in terms of the polylogarithm as follows [31]:

$$\bar{F}_{ss} = \frac{2n_0 \pi a^2}{x_0^2} \frac{k_B T}{x_s} \frac{Q(B)}{E_1(B)} [x_0 \ln(1 + De^{x_0}) - \text{Li}_2(-D) + \text{Li}_2(-De^{x_0})], \tag{42}$$

where $\text{Li}_2(z) = -\int_0^z t^{-1} \ln(1-t) \, dt$ is the dilogarithm [42]. Eq. (42) allows to analyze the variation of the frictional force with normal load analytically. Noticing that (here $F_{\text{tot}} = F_N + F_{\text{adh}}$)

$$\frac{a^2}{x_0^2} = \frac{a^2}{\left(\frac{3F_{\text{tot}}}{2\pi a^2}\right)^2 \left(\frac{V_a}{k_B T}\right)^2} = \left(\frac{k_B T}{V_a}\right)^2 \frac{4\pi^2 a^6}{9 F_{\text{tot}}^2} = \left(\frac{k_B T}{V_a}\right)^2 \frac{4\pi^2}{9 F_{\text{tot}}^2} \left(\frac{3R F_{\text{tot}}}{4E^*}\right)^2 = \left(\frac{k_B T}{V_a} \frac{\pi R}{2E^*}\right)^2, \tag{43}$$

is independent of $F_N$, thus we only need to focus on the load dependence of the following term:

$$I = x_0 \ln(1 + De^{x_0}) - \text{Li}_2(-D) + \text{Li}_2(-De^{x_0}), \tag{44}$$

in the r.h.s. of Eq. (42). Eq. (44) includes two leading parameters, $D$ and $De^{x_0}$, which depend on normal load, temperature and velocity. Its asymptotic behaviors for three limiting cases: (i) $D \ll 1$, $De^{x_0} \ll 1$; (ii) $D \ll 1$, $De^{x_0} \gg 1$; and (iii) $D \gg 1$ can be written as

$$\begin{cases} I \sim D(x_0 - 1)e^{x_0} + D, & \text{for } D \ll 1 \text{ and } De^{x_0} \ll 1 \\ I \sim \frac{1}{2}(x_0^2 - \ln^2 D) + D, & \text{for } D \ll 1 \text{ and } De^{x_0} \gg 1 \\ I \sim x_0 De^{x_0} + \frac{1}{2}\ln^2 D - \frac{1}{2}\ln^2 De^{x_0} \sim \frac{1}{2}x_0^2, & \text{for } D \gg 1 \end{cases} \quad (45)$$

Deriving Eq. (45), we used the asymptotic properties of dilogarithm [42]:

$$\text{Li}_2(-D) \sim \begin{cases} -D, & D \ll 1 \\ -\frac{1}{2}\ln^2(D), & D \gg 1 \end{cases} \quad (46)$$

Substituting the definitions of $x_0$ and $D$ into Eq. (45), the asymptotic equations for the friction force in different ranges of normal force read as:

$$\begin{cases} \bar{F}_{ss} \sim F_{ss}^0 D\left[\left(\left(\frac{F_N + F_{adh}}{F_0}\right)^{\frac{1}{3}} - 1\right)e^{\left(\frac{F_N + F_{adh}}{F_0}\right)^{\frac{1}{3}}} + 1\right] \sim \frac{F_{ss}^0 D}{2}\left[\left(\frac{F_N + F_{adh}}{F_0}\right) + \left(\frac{F_N + F_{adh}}{F_0}\right)^{\frac{2}{3}}\right], & \text{for } F_N + F_{adh} \ll F_0 |\ln D|^3, D \ll 1 \\ \bar{F}_{ss} \sim F_{ss}^0 \left[\frac{1}{2}\left(\left(\frac{F_N + F_{adh}}{F_0}\right)^{\frac{2}{3}} - \ln^2 D\right) + D\right] \sim \frac{F_{ss}^0}{2}\left(\frac{F_N + F_{adh}}{F_0}\right)^{\frac{2}{3}}, & \text{for } F_N + F_{adh} \gg F_0 |\ln D|^3, D \ll 1 \\ \bar{F}_{ss} \sim \frac{1}{2}x_0^2 \sim \frac{F_{ss}^0}{2}\left(\frac{F_N + F_{adh}}{F_0}\right)^{\frac{2}{3}}, & \text{for } D \gg 1 \end{cases}$$

$$(47)$$

where $F_{ss}^0 = 2\pi n_0 \left(\frac{k_B T}{V_a}\frac{\pi R}{2E^*}\right)^2 \frac{k_B T}{x_s}\frac{Q(B)}{E_1(B)}$, $F_0 = \frac{2\pi}{3}\left(\frac{k_B T}{V_a}\right)^3 \left(\frac{\pi R}{2E^*}\right)^2$.

Eq. (47) predicts that the variation of the friction force with normal load strongly depends on the value of parameter $D = r_{on}^0 Be^B E_1(B)$, which is a function of both the microscopic properties of contacts and the sliding velocity. To clarify this effect, we consider below different ranges of physical parameters, $\Delta E_{off}, \Delta E_{on}^0$ and $V$.

(i) For $\Delta E_{off} - \Delta E_{on}^0 \gg k_B T$, where $r_{on}^0 \gg 1$, the parameter $D$ can be larger or smaller than one depending on the value of $B$ (i.e., $V$). The value of $B = B_c$ separating two regimes can be deduced from the equation $r_{on}^0 B_c e^{B_c} E_1(B_c) = 1$. Since $r_{on}^0 \gg 1$, it's obvious that $B_c \ll 1$, the above equation can be approximated as $r_{on}^0 B_c \ln B_c = -1$, and its solution reads as

$$B_c = e^{W_{-1}(-1/r_{on}^0)} \approx \frac{1}{r_{on}^0 \ln r_{on}^0}, \quad (48)$$

where $W_{-1}(\cdot)$ is second real branch of the lambert $W$-function [43]. According to Eq. (8), the corresponding sliding velocity is

$$V_c = \frac{k_{off}^1}{e^{W_{-1}(-1/r_{on}^0)}}\frac{k_B T}{\kappa_b x_s} \approx k_{off}^1 \frac{k_B T}{\kappa_b x_s} r_{on}^0 \ln r_{on}^0, \quad (49)$$

Using Eqs. (47) - (49) we can make the following conclusions:

(a) In the low-velocity regime, $V \ll V_c$, where $B \gg B_c$, we get $D \gg 1$. In this regime for all values of the normal load the friction force scales as $F \propto F_N^{2/3}$, which is consistent with the perdition of Hertz contact (or DMT) theory.

(b) In the high-velocity regime, $V \gg V_c$, where $B \ll B_c$, we have $D \ll 1$. Eq. (47) show that in this regime the load dependence of friction changes from $F \propto F_N$ to $F \propto F_N^{2/3}$ as the normal load increases.

(ii) In the case of $\Delta E_{on}^0 - \Delta E_{off} > k_B T$, where $r_{on}^0 \ll 1$, it's easy to show that $D \ll 1$ since $0 \le Be^B E_1(B) \le 1$. Under these conditions Eq. (47) predicts two distinct scaling regimes in the friction-load curve, $F \propto F_N$ and $F \propto F_N^{2/3}$, that correspond to the ranges of low and high loads, respectively.

To illustrate the predictions of the above model, we choose the typical parameters for siloxane bonds [62]. **Fig. 8**a shows the normal load dependence of friction force for different sliding velocities. It is evident that the normal force dependence is switched between two regimes with a change in the sliding velocity. For the values of parameters used in **Fig. 8**, the value of critical velocity separating the two regimes is $V_c \approx 0.33$ μm/s. **Fig. 8**a shows that for low sliding velocities, $V < V_c$, the friction force increases with the normal load, following the power low, $F \propto F_N^{2/3}$ which is consistent with the results of kMC simulations [62]. However, for higher sliding velocity, $V \ge V_c$, the friction exhibits more complex behavior with $F \propto F_N$ in the low-load regime, and $F \propto F_N^{2/3}$ in the high-load regime. **Fig. 8**b-c show the velocity and temperature dependence of friction force for different sliding velocities. The shape of the curves is similar to that in **Fig. 2**.

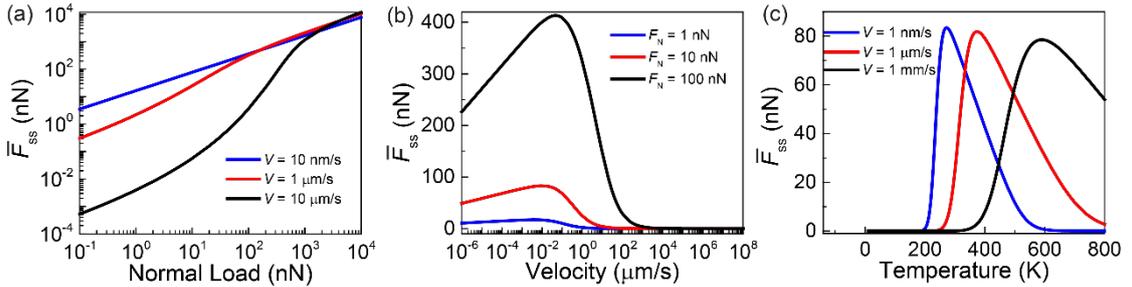

**Fig. 8**. Friction force as functions of (a) normal load, (b) sliding velocity and (c) temperature calculated with Eq. (42). The parameters used here: $\Delta E_{on}^0 = 0.6$ eV, $\Delta E_{off} = 1.1$ eV, $n_0 = 5/\text{nm}^2$, $V_a = 10$ Å$^3$; $k_b = 10$ N/m, $x_s = 0.2$ Å, $k_{on}^0 = 10^{13}$ s$^{-1}$, $k_{off}^0 = 10^{13}$ s$^{-1}$, $R_{tip} = 40$ nm, $E = 35$ GPa.

*3.6. Comparison between theory and experiments*

To check the validity of our model, we present in this section a comparison between our theoretical results and recent experimental data obtained for different interfaces under different conditions [14, 30, 31]. Noting that the available experimental data were obtained using FFM, the corresponding configuration should be considered, and Eq. (42) for the friction force should be used for the comparison with the experimental data. Based on the theoretical results and discussion presented in Sec. 3.4, we conclude that the novel temperature and velocity dependences observed in recent FFM experiments [14, 30, 31] indicate that at least two different types of contacts are formed at the considered frictional interfaces. To check whether our model can capture all observed behaviors, we use the following equation for the friction force measured in the FFM configuration with two types of contacts:

$$\overline{F}_{ss}^{tot} = \overline{F}_{ss}(F_N, V, T; n_0^I, V_a^I, E_{on}^I, E_{off}^I, \kappa_b^I) + \overline{F}_{ss}(F_N, V, T; n_0^{II}, V_a^{II}, E_{on}^{II}, E_{off}^{II}, \kappa_b^{II}), \quad (50)$$

where the both forces $\overline{F}_{ss}(F_N, V, T; n_0^I, V_a^I, E_{on}^I, E_{off}^I, \kappa_b^I)$ and $\overline{F}_{ss}(F_N, V, T; n_0^{II}, V_a^{II}, E_{on}^{II}, E_{off}^{II}, \kappa_b^{II})$ are described by Eq. (42). The superscripts I and II refer to the two types of contact, each of them is characterized by a set of parameters, which are tuned to fit the experimental data quantitatively. The equation (50) combines the concept of two-state distribution with Eq. (42) derived for the friction force in the FFM configuration. Discussions of physically motivated constraints on the fitting parameters and details of fitting procedures can be found in Section 3 of SM.

The comparison between the three sets of different experimental results [14, 30, 31] and our calculations using Eq. (50) is shown in **Fig. 9**. Overall, there is an excellent agreement between the theory and experiments. In particular, the non-linear normal load dependence of friction (**Fig. 9**a), the appearance of minimum in the velocity dependence of friction (**Fig. 9**b), which is accompanied by a minimum and maximum in the temperature dependence of friction (**Fig. 9**c) are nicely described by Eq. (50). In addition, the velocity weakening of friction is also well captured by the theory (**Fig. 9**a, d). Our consideration shows that velocity dependence of friction force strongly depends on the height of barriers for bond formation, and it can be significantly different for strong chemical bonds and weak van der Waals interactions. A good agreement between the model predictions and the experimental data obtained for various interfaces and environmental conditions suggests that our theoretical findings are robust and reliable.

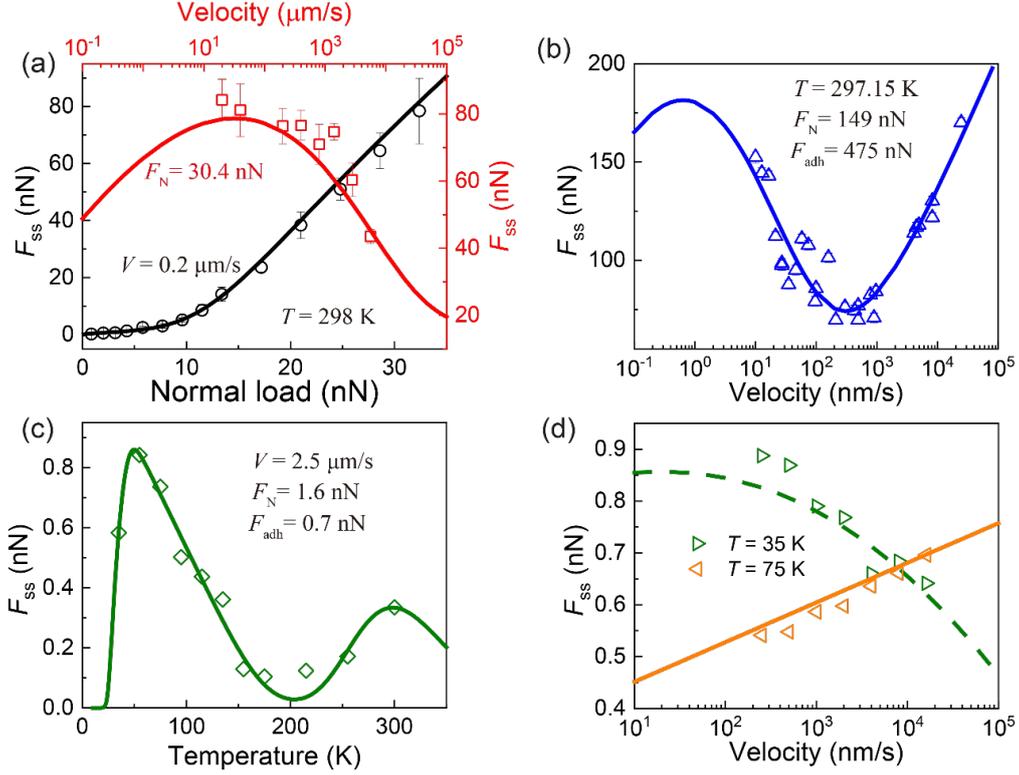

**Fig. 9.** Comparison between the theoretical predictions for the friction force and the experimental results obtained for various frictional interfaces. (a) Dependencies of friction force on normal load and velocity. Experimental data extracted from Fig. 1a-b of Ref. [31] (silicon tip/silicon substrate, $R_{\text{tip}} = 40$ nm, $E^* = 35.3$ GPa). (b) Velocity dependence of friction force. Experimental data extracted from Fig. 1a of Ref. [30] (silica tip/silica substrate, $R_{\text{tip}} = 125$ nm, $E^* = 35.3$ GPa). (c-d) Temperature and velocity dependence of friction force. Experimental data extracted from Fig. 1a and Fig. 3d of Ref. [14] (amorphous silicon oxide tip/Si(111) wafer, $R_{\text{tip}} = 20$ nm, $E^* = 35.3$ GPa), respectively. Points and lines represent the experimental data and fitting results of Eq. (50), respectively. Parameters in (a): $E_{\text{on}}^{\text{I}} = 0.87$ eV, $E_{\text{off}}^{\text{I}} = 0.78$ eV, $n^{\text{I}} = 7.92$ nm$^{-2}$, $V_a^{\text{I}} = 43.2$ Å$^3$; $E_{\text{on}}^{\text{II}} = 0.04$ eV, $E_{\text{off}}^{\text{II}} = 0.58$ eV, $n^{\text{II}} = 1$ nm$^{-2}$, $V_a^{\text{II}} = 24.5$ Å$^3$. Parameters in (b): $E_{\text{on}}^{\text{I}} = 0.889$ eV, $E_{\text{off}}^{\text{I}} = 0.796$ eV, $n^{\text{I}} = 1.76$ nm$^{-2}$, $V_a^{\text{I}} = 25.4$ Å$^3$; $E_{\text{on}}^{\text{II}} = 0.05$ eV, $E_{\text{off}}^{\text{II}} = 0.54$ eV, $n^{\text{II}} = 1.14$ nm$^{-2}$, $V_a^{\text{II}} = 50$ Å$^3$. Parameters in (c-d): $E_{\text{on}}^{\text{I}} = 0.75$ eV, $E_{\text{off}}^{\text{I}} = 0.41$ eV, $n^{\text{I}} = 4.1$ nm$^{-2}$, $V_a^{\text{I}} = 59.7$ Å$^3$; $E_{\text{on}}^{\text{II}} = 0.08$ eV, $E_{\text{off}}^{\text{II}} = 0.23$ eV, $n^{\text{II}} = 0.66$ nm$^{-2}$, $V_a^{\text{II}} = 5.8$ Å$^3$. Other parameters are the Same for all the figures: $k_{\text{on}}^0 = k_{\text{off}}^0 = 10^{13}$ Hz, $x_s = 0.2$ nm, $\kappa_b^{\text{I}} = 10$ N/m and $\kappa_b^{\text{II}} = 1$ N/m.

## 4. Conclusion

We developed an analytical model for description of friction mediated by dynamical formation and rupture of microscopic interfacial contacts. The model accounts for the presence of various types of contacts at the frictional interface and predicts novel frictional behaviors, which are amenable to experimental observations. The main findings are summarized as follows.

i) We derived analytical equations, describing dependencies of steady-state friction force on sliding velocity, normal load and temperature on the time and length scales that are relevant to tribological experimental conditions. Our analytical calculations are in excellent agreement with the experimental data obtained for various frictional interfaces and with the results of stochastic simulations performed for the multi-contact model.

ii) In the range of velocities, normal loads and temperatures typical for friction measurement at the nanoscale, our model predicts the velocity-temperature scaling relationship, which relies on the interplay between the effects of shear and temperature on the rupture of interfacial contacts. The proposed scaling can be used to extrapolate the simulation results to a range of very low sliding velocities used in nanoscale friction experiments, which is still unreachable by simulations. The predicted velocity regimes of friction are confirmed by recent experimental and theoretical findings.

iii) We derived analytical equations for the friction forces at interfaces characterized by three typical types of distributions of contact properties. The distributions considered here can be used to approximate any realistic distribution of contact properties measured experimentally or obtained in simulations

iv) Considering FFM configuration, we derived an analytical equation describing the dependence of friction force on sliding velocity, normal load and temperature. The predicted nonlinear dependence of friction on normal load differs significantly from the predictions of the commonly used Hertz and Derjaguin-Müller-Toporov (DMT) models, and is confirmed by a recent FFM experiment [31].

v) For the first time, we predicted double-peak dependencies of friction on temperature and velocity, which are amenable to experimental observations Temperature, velocity and load dependencies of friction measured in recent FFM experiments for various frictional interfaces [14, 30, 31] can be well explained by our theoretical model.

The analytical solutions derived in this work provide a framework for the interpretation of the experimental data on friction at interfaces including microscopic contacts and open new pathways for the rational control of frictional response.

## Acknowledgments

W.O. acknowledges the financial support from the Planning and Budgeting Committee fellowship program for outstanding postdoctoral researchers from China and India in Israeli Universities, from the Ratner Center for Single Molecule Science at Tel Aviv University and from the National Natural Science Foundation of China (No. 11890673). M. M. acknowledges the financial support from the Thousand Young Talents Program and the NSFC grant No. 11890673, 11632009, and 11772168. M.U. acknowledges the financial support of the Israel Science Foundation, Grant No. 1141/18, and of the Deutsche Forschungsgemeinschaft (DFG), Grant No. BA 1008/21.

## Supplementary material

Supplementary material associated with this article can be found in the online version.

## Author contributions

M.U. conceived the original idea behind this study. W.O. derived all the equations, wrote the MATLAB codes for numerical calculations and fitting of analytical equations and analyzed experimental data. W.O. and M.U. analyzed the theoretical results with contribution from M.M. M.M. wrote the FORTRAN codes for the stochastic simulations and Y.C. performed the simulations. M.M. and Y.C. analyzed the simulations data with contributions from W.O. and M.U. W.O. wrote the initial draft and all authors contributed to the writing of this manuscript.

# Supplementary material for "Load-velocity-temperature relationship in frictional response of microscopic contacts"

*Wengen Ouyang[1], Yao Cheng[2], Ming Ma[2,\*], Michael Urbakh[1,\*]*

[1] School of Chemistry and The Sackler Center for Computational Molecular and Materials Science, Tel Aviv University, Tel Aviv 69978, Israel

[2] State Key Laboratory of Tribology, Department of Mechanical Engineering and Center for Nano and Micro Mechanics, Tsinghua University, Beijing 100084, People's Republic of China

[\*]Corresponding author.
E-mail address: urbakh@tauex.tau.ac.il (M. Urbakh).
E-mail address: maming16@tsinghua.edu.cn (M. Ma)

# 1 Numerical simulation

To prove the accuracy of the analytical expression given by Eq. (11) in the main text, we performed stochastic simulations of the multi-contact model for a single type of contacts, as that presented in [1, 2].

Different from the protocol used in [1, 2], here we consider a slider driven at a constant velocity, $V_{dr}$, rather than attached through a spring to a stage, which moves with constant velocity. Using this configuration allows a direct comparison with the results of theoretical model. We checked that both driving protocols give similar time-averaged friction forces in a broad range of velocities and temperatures.

For the slider driven at constant velocity, the external force (friction force) acted on it is calculated as:

$$F_{ext} = \sum_{i=1}^{N} q_i f_i^x. \tag{S1}$$

In Eq. (S1), $f_i^x$ is the force between the slider and the $i^{th}$ bound contact. The parameter $q_i$ characterizes the state of an individual contact: $q_i = 1$ corresponds to a bound state, while $q_i = 0$ corresponds to a ruptured bond (unbound state). Except of rates of contact formation and rupture ($k_{on}^i$ and $k_{off}^i$), another random variable $\xi_i$ is introduced to calculate the state of contact, $q_i$, at each time step. Time evolution of the state parameter, $q_i$ is calculated as [2]:

$$q_i(t + \Delta t) = q_i(t) - q_i(t)\theta(\Delta t k_{off}^i - \xi_i) + [1 - q_i(t)]\theta(\Delta t k_{on}^i - \xi_i), \tag{S2}$$

where $\theta(z)$ is a Heaviside step function, ensuring the contact $i$ attaches to the slider from the unbound state when $\Delta t k_{on}^i > \xi_i$, and ruptures from the bound state when $\Delta t k_{off}^i > \xi_i$.

The rate for contact rupture $k_{off}^i$, is influenced by the contact elongation, $l$, at this moment. When the instantaneous elongation $\Delta l_i$ is large enough, the potential barrier for bond rupture $\Delta E_{off}$ vanishes as follows:

$$k_{off}^i = k_{off}^0 \exp[(-(\Delta E_{off} - \kappa_b x_s \Delta l_i)/k_B T)]. \tag{S3}$$

The rate for bond formation is described as

$$k_{on} = k_{on}^0 \exp(-\Delta E_{on}/k_B T). \tag{S4}$$

**Fig. S1** illustrates the comparison between theoretical and numerical results for velocity and temperature dependences of steady state friction. It shows excellent agreement between the results of analytical model and numerical simulation, thus proving the validity of the theoretical equation.

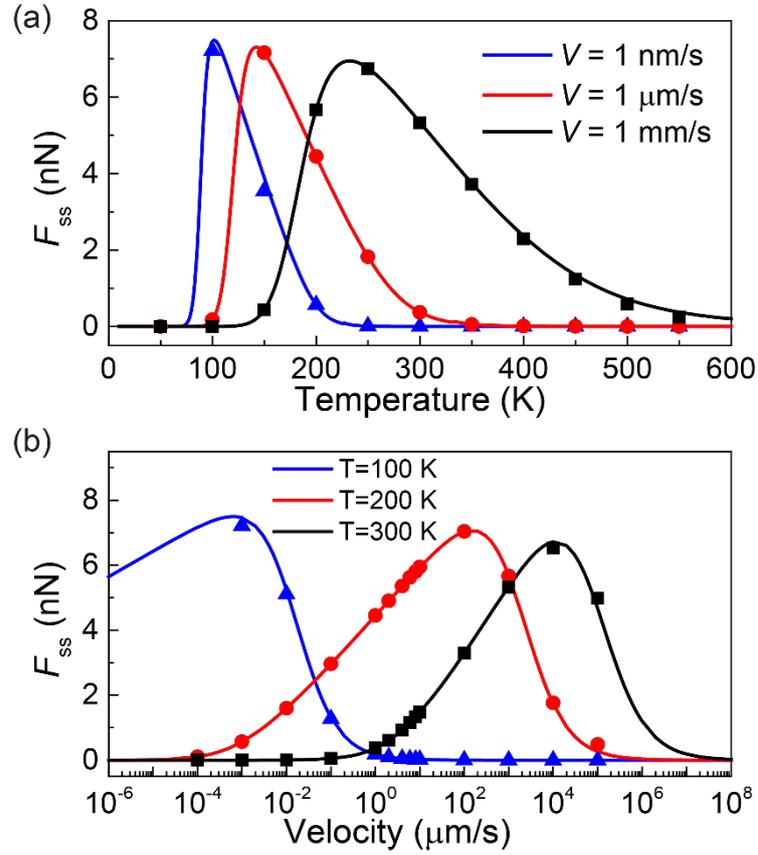

**Fig. S1.** Velocity and temperature dependences of friction force (a) Temperature dependence of friction for different sliding velocity. (b) Velocity dependence of friction for different temperature. The solid lines are calculated using Eq. (11) in the main text and the points are calculated using stochastic numerical simulation. The parameters used are as follows: $k_{on}^0 = k_{off}^0 = 10^{13}$ s$^{-1}$, $\Delta E_{off} = 0.40$ eV, $\Delta E_{on} = 0.20$ eV, $\kappa_b = 10$ N/m, $x_s = 0.2$ nm, $N = 100$.

## 2 Temperature and velocity dependence of friction for exponential distribution

The Gamma distribution discussed in the main text reduces to the exponential distribution when $k = 1$ (see Fig. 3b in the main text). Fig. 6 in the main text shows the temperature and velocity dependences of friction for Gamma distribution with $k = 2$. Here we show the results obtained for $k = 1$ using the same parameters.

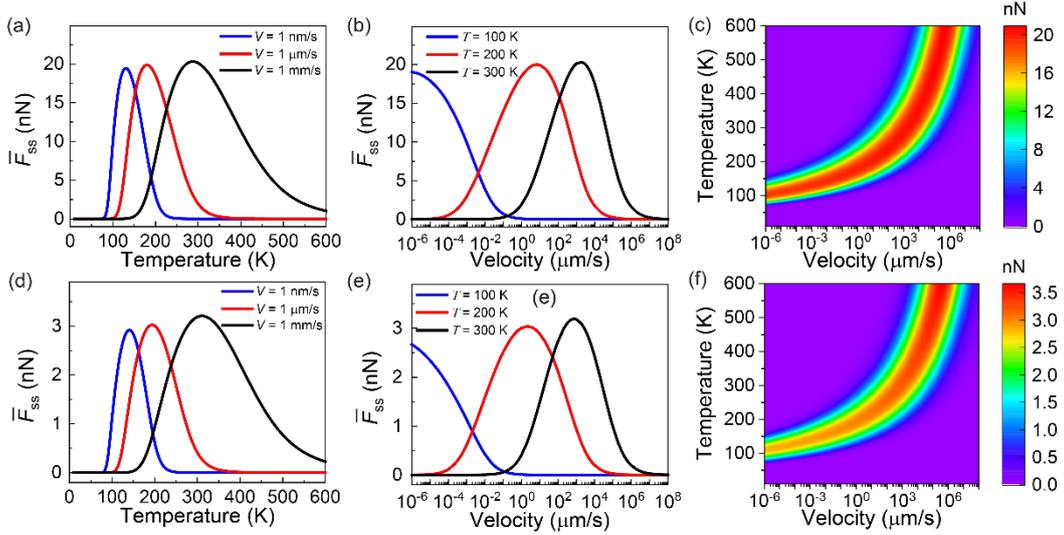

**Fig. S2.** Friction force as a function of temperature and velocity calculated for gamma PDFs with $k = 1$ and two different widths. The temperature (a, d) and velocity (b, e) dependences and their contour maps (c, f) of steady-state friction for $W = 0.1$ eV and $W = 1$ eV, respectively. Parameters used here are the same as that in Fig. 6 in the main text.

**Fig. S2** illustrates the temperature and velocity dependencies of friction for the exponential distribution with the same parameters as in Fig. 6 in the main text. One can see that the qualitative behavior of the friction force as a function of temperature and velocity is similar to that illustrated in Fig. 6. The difference is that the positions of the friction peaks shift to higher temperatures and lower velocities with increasing the value of $k$, and the magnitude of the friction force reduces significantly with increasing $k$.

# 3  Constraints on the fitting parameters and the fitting procedure

The model includes the following parameters for each set of contacts: $n_0, V_a, E_{on}, E_{off}, k_{on}^0, k_{off}^0, \kappa_b, x_s$. In order to reduce the arbitrariness in their choice, we took into account physically motivated constraints on their values, as described in our recent work [3]. Following the kinetic Monte Carlo (kMC) simulations [4, 5], the values of the attempt frequency are chosen as $k_{on}^0 = k_{off}^0 = 10^{13}$ Hz. Following the previous theoretical studies of single molecule force spectroscopy [6, 7], the $x_s$ is chosen as 0.2 nm. It's more difficult to estimate effective stiffness of the elastic springs characterizing bonds. In this Letter we considered two types of bonds:

## 3.1  Effective stiffness for strong interaction

For the strong bonds (e.g., chemical bonding), considering the siloxane bond (Si-O-Si) as an example, we used the results of ab initio simulations of a silicate molecule $H_6Si_2O_7$ [8, 9] to estimate the effective bond stiffness. The Si-O-Si is characterized by both stretching and bending stiffnesses, whose values are provided by the ab initio simulations. Then, using the methods developed in Ref. [10], we found that the effective stiffness $\kappa_b$ ranging from 10 N/m to 20 N/m. In our calculations, we used $\kappa_b$=10 N/m, and we checked that the friction force is not sensitive to changes of $\kappa_b$ in the above range.

## 3.2  Effective stiffness for weak interaction

Weak (van der Waals) bonds, can be described by Lennard-Jones (LJ) potential. In this case, using typical LJ parameters found for various materials [11], $\kappa_b$ is estimated as ~0.8-1.2 N/m. According to this, we used $\kappa_b$=1 N/m to represent weak contacts formed due to the van der Waals interaction.

## 3.3  The fitting procedure

The above consideration provides the values of four parameters: $k_{on}^0, k_{off}^0, \kappa_b, x_s$. There are additional four parameters ($n_0, V_a, E_{on}, E_{off}$) that can be tuned to fit the experimental data. The physical range of $E_{on}, E_{off}$ for strong Si-O-Si bond has been estimated previously [4, 5, 12], giving $E_{on} \sim 0.6 - 1.4$ eV and $E_{off} \sim 0.7 - 1.3$ eV. For weak contacts, Barel *et al.* [1] proposed much smaller values: $E_{on} \sim 0.05$ eV and $E_{off} \sim 0.15$ eV. According to previous studies [13-15], the activation volume $V_a$ ranges from 4 Å³ to 268 Å³. The number of binding sites per unit area for Si-O-Si bond on a fully hydroxylated silica surface, $n_0$, which value can be estimated experimentally, ranges from 4/nm² to 10/nm² [16-18].

Considering two types of contacts, eight parameters, ($n_0^I, V_a^I, E_{on}^I, E_{off}^I$) and ($n_0^{II}, V_a^{II}, E_{on}^{II}, E_{off}^{II}$), are tuned to fit the experimental data in Fig. 9 in the main text. Following the above considerations, in the fitting procedure the range of parameters for strong and weak contacts were constrained as follows: $0.1/\text{nm}^2 \leq n_0^I \leq 10/\text{nm}^2$, $1\ \text{Å}^3 \leq V_a^I \leq 100\ \text{Å}^3$, $0.1\ \text{eV} \leq E_{on}^I \leq 1.5\ \text{eV}$, $0.1\ \text{eV} \leq E_{off}^I \leq 1.5\ \text{eV}$ and $0.1/\text{nm}^2 \leq n_0^{II} \leq 10/\text{nm}^2$, $1\ \text{Å}^3 \leq V_a^{II} \leq 100\ \text{Å}^3$, $0.01\ \text{eV} \leq E_{on}^{II} \leq 0.5\ \text{eV}$, $0.1\ \text{eV} \leq E_{off}^{II} \leq 1.0\ \text{eV}$.

It can be seen form Fig. 9 in the main text that the fitted value of energy barriers for the strong contacts ($E_{on}^I, E_{off}^I$) extracted from different experimental data agree well with the above estimation. The similarity between the estimated activation energy barriers by fitting the various experimental data with Eq. (50) in the main text prove the robustness of the model and convergence of the fitting procedure used here.

## 4   Effect of humidity on the fitting results of Fig. 9b

In Fig. 9b, we show the comparison between the fitted velocity dependence of friction with our theoretical model and the experimental results, where we assumed the kinetic parameters of two types of contacts are independent. However, according to the explanation proposed by Tian et al. [19], the energy barrier for breaking a contact and the stiffness of the contact should be identical since they form the same type of bond although their activation energy barriers are different. To check its effect on the fitting results, we refitted the experimental data in Fig. 9b in the main text with the following constraints: $E_{off}^I = E_{off}^{II}$ and $\kappa_b^I = \kappa_b^{II} = 10$ N/m. The fitted results are illustrated in **Fig. S3**a, we can see from this figure that the agreement between the new fitting results and experiment is worse than that of Fig. 9b in the main text. We attribute the reason of it to the effect of water between two surfaces since the relative humidity (RH) is 45 % in this measurement. In this case, the dangling bond Si-O⁻ is easily passivated by the water and becomes to Si-OH, thus the first type of contacts is Si-OH and the second type of contacts is dominated by the interfacial water instead of the dangling bond Si-O⁻.

To further prove our point, we fitted another experimental results measured at low humidity (RH < 1 %), which is presented in Fig. S4a in the supplementary information of Ref. [19]. In this case, the fitted velocity dependence of friction with the same constraints gives very good agreement with the experimental data (see **Fig. S3**b), which indicates that the two dominated contacts at the interfaces are Si-OH and Si-O⁻ in this case, consistent with the explanation proposed by Tian et al. [19].

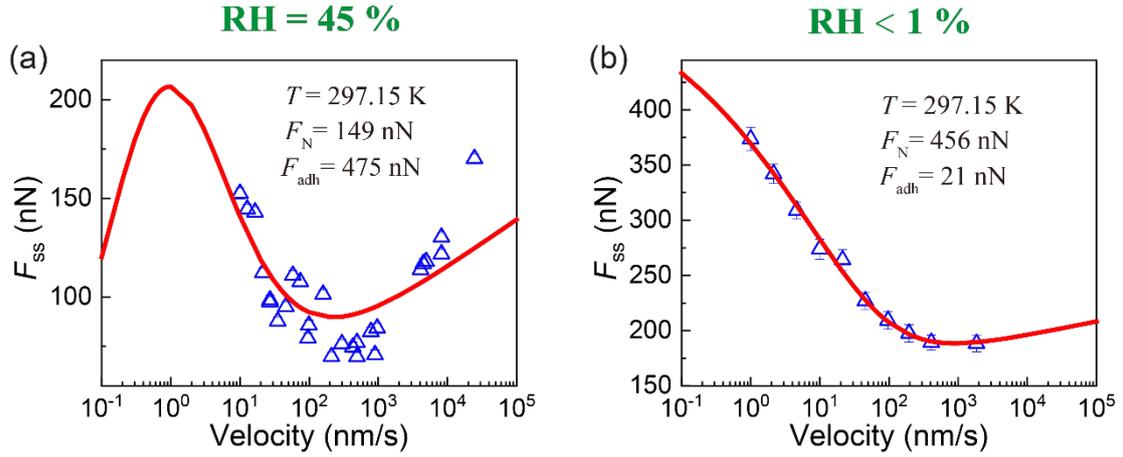

**Fig. S3**. Comparison between the theoretical predictions for the friction force and the experimental results of velocity dependence of friction force. The experimental data (blue triangles) are extracted from (a) Fig. 1 and (b) from Fig. S4a of Ref. [19] (silica tip/silica substrate, $R_{\text{tip}} = 125$ nm, $E^* = 35.3$ GPa). The red lines are the fitting results. Parameters in (a): $E^{\text{I}}_{\text{on}} = 0.654$ eV, $E^{\text{I}}_{\text{off}} = 0.659$ eV, $n^{\text{I}} = 10$ nm$^{-2}$, $V^{\text{I}}_a = 1$ Å$^3$; $E^{\text{II}}_{\text{on}} = 0.049$ eV, $E^{\text{II}}_{\text{off}} = 0.659$ eV, $n^{\text{II}} = 0.378$ nm$^{-2}$, $V^{\text{II}}_a = 51$ Å$^3$. Parameters in (b): $E^{\text{I}}_{\text{on}} = 1.2994$ eV, $E^{\text{I}}_{\text{off}} = 1.2996$ eV, $n^{\text{I}} = 1.32$ nm$^{-2}$, $V^{\text{I}}_a = 58.8$ Å$^3$; $E^{\text{II}}_{\text{on}} = 0.05$ eV, $E^{\text{II}}_{\text{off}} = 1.2996$ eV, $n^{\text{II}} = 0.233$ nm$^{-2}$, $V^{\text{II}}_a = 50.5$ Å$^3$. Other parameters are the Same for all the figures: $k^0_{\text{on}} = k^0_{\text{off}} = 10^{13}$ Hz, $x_{\text{s}} = 0.2$ nm, $\kappa^{\text{I}}_{\text{b}} = \kappa^{\text{II}}_{\text{b}} = 10$ N/m.

In short summary, the chemistry of the contact is sensitive to the environmental humidity for the silica interfaces. Further experiments are required to identify the second type of contact.